\documentclass[preprint,review,12pt]{elsarticle}

\usepackage{amssymb}

\journal{Journal of Magnetism and Magnetic Materials}

\begin{document}

\begin{frontmatter}



\title{Terahertz magnetoelectric response via electromagnons in magnetic oxides}


\author[label1]{N. Kida}
\author[label2,label3,label4]{Y. Tokura}
\address[label1]{Department of Advanced Materials Science, The University of Tokyo, 5-1-5 Kashiwa-no-ha, Kashiwa 277-8561, Japan}
\address[label2]{Multiferroics Project (MF), ERATO, Japan Science and Technology Agency (JST), c/o Department of Applied Physics, The University of Tokyo, 7-3-1 Hongo, Bunkyo-ku, Tokyo 113-8656, Japan}
\address[label3]{Department of Applied Physics, The University of Tokyo, 7-3-1 Hongo, Bunkyo-ku, Tokyo 113-8656, Japan}
\address[label4]{Cross-Correlated Materials Research Group (CMRG) and Correlated Electron Research Group (CERG), ASI, RIKEN, 2-1 Hirosawa, Wako, 351-0198, Japan}

\begin{abstract}
Terahertz magnetic resonance driven by the light electric field, now referred to as electromagnon, has stimulated interest due to its strong candidate for future spin-electronics. One unique characteristic of electromagnons is the terahertz magnetochromism, a color change by external magnetic field, as recently demonstrated in the hexaferrite. By taking perovskite manganites and hexaferrite as model cases, the current understating  of the electromagnon activity is discussed in terms of the symmetric exchange mechanism. 
\end{abstract}

\end{frontmatter}


\section{Introduction}
The control of the magnetic state by external electric field or the polarization state by external magnetic field is a demanding functionality for spin-electronics technology. However, the materials showing such magnetoelectric (ME) effect are still rare. Therefore, recent findings of the gigantic ME response in multiferroics endowed with ferroelectric and magnetic orders have stimulated considerable interest \cite{YTokuraREV,TArima}. 
To explain the magnetically-induced ferroelectric polarization $P_{\rm s}$ in multiferroics, the spin-current \cite{HKatsura} or inverse Dzyaloshinskii-Moriya (DM) model \cite{IASergienko} is proposed, which is simply expressed by
\begin{equation}
P_{\rm s}\propto e_{ij}\times (S_i\times S_j),
\end{equation}
where $e_{ij}$ the unit vector connecting the neighboring spins $S_i$ and $S_j$ [Fig. 1(a)] \cite{HKatsura,IASergienko,Mostovoy}. In this model, $P_{\rm s}$ can be produced along the direction perpendicular to the modulation vector $q$ and within the spiral spin plane [Fig. 1(a)]. This is consistent with the experimental data of non-collinear spiral magnets such as perovskite $R$MnO$_3$ ($R$ represents the rare-earth ions), in which $P_{\rm s}$ appears along $c$-axis in the $bc$ spiral spin ordered phase (see, the right panel of Fig. 2) \cite{TKimura}.

An important consequence of such ME effect in multiferroics is the dynamical coupling between ferroelectricity and magnetism \cite{YTokuraNKida}.  One such example is the electric-dipole active magnetic resonance, now termed electromagnon. Usually, the magnetic resonance driven by the light $H$ vector shows up in the magnetic permeability spectrum $\mu(\omega)$ at gigahertz to terahertz frequencies. This is well known as an antiferromagnetic resonance (AFMR) for the case of antiferromagnets. On the other hand, the electromagnon can be excited by the light $E$ vector rather than by the light $H$ vector and thus the resonance shows up in the dielectric constant spectrum $\epsilon(\omega)$. 
Up to date, the emergence of electromagnons was argued in a variety of magnetic oxides such as perovskite $R$MnO$_3$ \cite{PimenovREV,DSenffREV,NKidaRMnO3review}, $R$Mn$_2$O$_5$ \cite{ABSushkov}, hexagonal YMnO$_3$ \cite{SPailhes}, BiFeO$_3$ \cite{MCazayous}, Ba$_2$Mg$_2$Fe$_{12}$O$_{22}$ \cite{NKidaYpart1,NKidaYpart2}, CuFe$_{1-x}$Ga$_x$O$_2$ \cite{SSeki}, Ba$_2$CoGe$_2$O$_7$ \cite{IKezsmarki}, and Dy$_3$Fe$_5$O$_{12}$ garnet \cite{PDRogers}. Contrary to the origin of the ferroelectricity in multiferroics based on Eq. (1), the electromagnon activity mainly comes from the symmetric exchange interaction, as discussed in $R$MnO$_3$ as a model case \cite{RAguilar,SMiyahara,MPVStenberg,MMochizuki_EM}. This clearly means that the fingerprint of the electromagnon would be generally found in a variety of magnets, not limited to multiferroics.

\section{Electromagnons in $R$MnO$_3$}
First possible signature of the electromagnon was identified as a single peak around 2 meV (4 meV $\approx$ 1 THz) in $\epsilon$ spectrum in prototypical multiferroics, TbMnO$_3$ and GdMnO$_3$ \cite{APimenov1}. As the conventional AFMR is observed at terahertz frequencies in perovskite $R$MnO$_3$ \cite{AAMukhin}, the measurements of the complete set of the light-polarization dependence with respect to the crystallographic axis is indispensable to identify the contribution of $\mu(\omega)$ to the optical constants $\tilde{n}$ (=$\sqrt{\epsilon\mu})$, as carefully performed in DyMnO$_3$ \cite{NKidaDyMnO3} and TbMnO$_3$ \cite{YTakahashi}. As an example of the electromagnon in $R$MnO$_3$, we show in Fig. 2 the imaginary part of $\epsilon\mu$ spectra of DyMnO$_3$ for $E^\omega\parallel a$ and $H^\omega\parallel c$ at various spin ordered phases, measured by using terahertz time-domain spectroscopy \cite{NKidaRMnO3review,NKidaDyMnO3}. The quantity $\epsilon\mu$ is used as there is a contribution of $\mu$ of AFMR for $H^\omega\parallel c$ in this configuration. As lowering temperature, the electromagnon for $E^\omega\parallel a$ is discerned in the sinusoidal collinear spin ordered phase below 39 K and finally grows in intensity in the thermally-induced $bc$ spiral spin ordered phase below 19 K. Among $R$MnO$_3$, DyMnO$_3$ produces the largest spectral weight of the electromagnon. Contrary to early experiment, the electromagnon spectrum is revealed to spread over an energy range of 1--10 meV and shows the two peak structures around 2 meV and 6 meV (Fig. 2).

In an early stage of the electromagnon study, the origin of the electromagnon for $E^\omega\parallel a$ was believed as a result of the rotation of the spiral spin plane, and hence the modulation of $P_{\rm s}$ based on Eq. (1). However, the current understanding is that the exchange striction, inherent to the chemical lattice of the perovskite structure, can act as a source of the electromagnon activity. A direct experimental proof of above consideration was provided from the measurements of  the effect of the magnetic field on the electromagnon spectra of DyMnO$_3$ \cite{NKidaDyMnO3}. At 7 K in magnetic fields above 2 T, we can induce the flop of the spiral spin plane from $bc$ to $ab$ \cite{TKimura}, as schematically shown in right panels of Fig. 2. As seen in Fig. 2, the electromagnon for $E^\omega\parallel a$ is clearly visible even in the magnetically-induced $ab$ spiral spin ordered phase, as exemplified by $\epsilon\mu$ spectra at 3.3 T and 5.9 T; this is a direct proof that the electromagnon is independent of the orientation of the spiral spin plane and uniquely shows up only along $a$-axis. The same tendency was lately confirmed in the thermally-induced $ab$ spiral spin ordered phase of Gd$_{0.7}$Tb$_{0.3}$MnO$_3$ with the complete set of the light-polarization dependence \cite{NKidaGdTbMnO3} and the magnetically-induced $ab$ spiral spin ordered phase of TbMnO$_3$ \cite{RAguilar}.

To explain the unique selection rule of the observed electromagnons, i.e., only electric-dipole active along $a$-axis, the symmetric exchange mechanism is proposed \cite{RAguilar,SMiyahara}, in which the non-collinear spin structure can produce the polarization $\Delta P_{ij}$, as simply expressed by
\begin{equation}
\Delta P_{ij}\propto \Delta S_i\cdot S_j.
\end{equation}
For the case of $R$MnO$_3$, $\Delta P_{ij}$ is built in via the Mn-O-Mn bond distribution \cite{RAguilar} or staggered $e_g$ orbital ordered state along $a$-axis \cite{SMiyahara}, as schematically shown in Fig. 1(b). In $E^\omega\parallel a$ configuration, $\Delta P_{ij}$ is only produced along $a$-axis as $\Delta S_i$ is not perpendicular to $S_j$. Indeed, the calculation based on above scenario can reproduce the higher-lying electromagnon around 6--10 meV \cite{RAguilar,SMiyahara,MPVStenberg}. The validity of this scenario was confirmed by the systematic investigation on the electromagnon spectra by changing the ionic radius of $R$ ($R$ = Gd, Gd$_{0.7}$Tb$_{0.3}$, Gd$_{0.5}$Tb$_{0.5}$, Gd$_{0.1}$Tb$_{0.9}$, Tb$_{0.41}$Dy$_{0.59}$, and Dy) and hence the exchange energy $J$ \cite{JSLee}. Recently, the lower-lying electromagnon around 2 meV was also reproduced by introducing the deformation of the higher order spiral spin plane \cite{MMochizuki_EM}. Therefore, the electromagnon of $R$MnO$_3$ appeared in the energy range of 1--10 meV can be regarded as the electric-dipole active one magnon excitation. Contrary to the origin of the ferroelectricity in $R$MnO$_3$ as expressed by Eq. (1), the electromagnon activity is induced via the symmetric exchange mechanism based on Eq. (2).

\section{Terahertz magnetoelectric response via electromagnons}
Although the electromagnon is now widely observed in a variety of antiferromagnets, there is a strong demand for the electromagnon in ferromagnets. One such candidate is a Y-type hexaferrite Ba$_2$Mg$_2$Fe$_{12}$O$_{22}$. At room temperature, Ba$_2$Mg$_2$Fe$_{12}$O$_{22}$ shows the ferrimagnetic order within the (001) plane, composed of two magnetic sublattice blocks $L$ and $S$ [Fig. 3(c)]. It undergoes the proper screw spin transition around 195 K at which spins rotate with the propagation vector along [001]. With decreasing temperature below 50 K, spins decline along [001] and the longitudinal conical spin structure characterized by the conical angle $\theta$ is finally formed, as schematically shown in Fig. 3(c). At 6 K in zero magnetic field, $\theta$ was estimated to be about 20$^\circ$ according to the recent neutron scattering experiments \cite{SIshiwataNeutron}.

A remarkable characteristic of Ba$_2$Mg$_2$Fe$_{12}$O$_{22}$ is the emergence of the ferroelectricity \cite{SIshiwata,KTaniguchi}. In the magnetic field along [100], the transverse conical spin ordered phase can be realized as a result of declining the cone axis toward [100] \cite{SIshiwata,HSagayama}. It can be viewed as the spiral plus ferromagnetic orders; therefore, according to Eq. (1), $P_{\rm s}$ emerges along [120] both perpendicular to the directions of the modulation vector [001] and the magnetic field [100] [see, Fig. 3(c)], which is confirmed experimentally \cite{SIshiwata,KTaniguchi}. Furthermore, Ba$_2$Mg$_2$Fe$_{12}$O$_{22}$ is unique in the sense that the thermally-induced spin structure can be easily controlled by applying the magnetic field along [001]; in this case, although there is no route to induce the ferroelectricity [Fig. 1(c)], the spin structure can transform from the proper screw to the ferrimagnetic through the conical spin ordered phases.

In the longitudinal conical spin ordered phase realized in zero magnetic field [Fig. 3(c)], the electromagnon was identified on the basis of the contemporary measurements of terahertz time-domain spectroscopy and inelastic neutron scattering \cite{NKidaYpart1}. Figure 3(a) shows the light-polarization dependence of $\epsilon_2$ spectra, measured at 5 K. In this case, the contribution of $\mu(\omega)$ to $\tilde{n}$ is negligible, hence $\epsilon_2$ was used as a quantity. The sharp resonance absorption is discerned around 2.8 meV when $E^\omega$ was set parallel to [001]. By changing the direction of $E^\omega$ from [001] to [120], while keeping the direction of $H^\omega$ along [100], this resonance disappears. Therefore, it was assigned to the electric-dipole active mode, inherent to the longitudinal conical spin structure. Accordingly, the clear peak structure is observed around 3 meV in the magnetic excitation spectrum at the zone center $\delta=0$ [$Q=(2+\delta, -2-\delta, 6)$ in the Brillouin zone], as shown in Fig. 3(b). The magnon dispersion curve is parabolic as a function of $\delta$ [inset of Fig. 3(b)]. Based on two different experimental probes, the observed resonance around 2.8 meV is assigned to the electromagnon, the first example of the electromagnons in the ferro(ferri)magnets.

As mentioned, the external magnetic field can dramatically modify the spin structure of Ba$_2$Mg$_2$Fe$_{12}$O$_{22}$. By an application of the magnetic field along [001], the magnetization yields the large saturation moment $\sim8$ $\mu_{\rm B}$ per formula unit at 7 T, and correspondingly $\theta$ can be perfectly controlled from $0^\circ$ (proper screw) to $90^\circ$ (ferrimagnetic) through $0^\circ<\theta<90^\circ$ (longitudinal conical), as schematically shown in the right panel of Fig. 4. This magnetic control of the spin structures leads to the change of the spectral shape of the electromagnon inherent to the longitudinal conical spin ordered phase. Near the proper screw to the longitudinal conical spin transition temperature at 54 K in zero magnetic field, no signature of the electromagnon is found (Fig. 4). On the contrary, in magnetic fields, there is a remarkable signature of the electromagnon due to the evolution of the longitudinal conical spin order. This is viewed as a gigantic terahertz magnetochromism \cite{NKidaYpart2}. The gigantic change of $\epsilon_2$ is identified even at 0.3 T. This is because that the small magnetic field is enough to induce the longitudinal conical spin structure. With further increasing magnetic fields, the spins become collinear and exhibit ferrimagnetic order along [001]. In this phase, the electromagnon is diminished. 

The mechanism of the electromagnon activity may share the same origin of $R$MnO$_3$. In Ba$_2$Mg$_2$Fe$_{12}$O$_{22}$, the longitudinal conical spin structure is composed of $L$ and $S$ blocks of Fe ions along [001] [Fig. 3(c)]. We focus on three Fe ions located between $L$ and $S$ blocks on the basis of the neutron scattering experiment \cite{NMomozawa}. The spin component of these Fe ions projected on (120) plane is schematically shown in Fig. 1(d). In this case, non-zero component of $\Delta P_{ij}$ along [001] is induced in response to $E^\omega$ as $\Delta S_i$ is not perpendicular to $S_j$. As the magnitude of $\Delta P_{ij}$ depends on $\theta$, the electromagnon intensity should be scaled with $\sin^2\theta$ and shows the maximum at $\theta=45^{\circ}$. The validity of which can be identified by the detailed magnetic field effect of the electromagnon; the electromagnon intensity indeed reaches the maximum around $\theta=45^{\circ}$ \cite{NKidaYpart2}.

\section{Summary and Prospect}
In cases of perovskite $R$MnO$_3$ and Ba$_2$Mg$_2$Fe$_{12}$O$_{22}$, the mechanism of the electromagnon activity is likely from the exchange striction induced by the non-collinear spin order. This is in contrast to the origin of the ferroelectricity in these compounds induced by the spiral spin order.
Therefore, we can anticipate that the electromagnon activity is ubiquitous and its signature should be detected in a variety of magnets, not restricted to multiferroics. 

To identify the electromagnon activity, the light-polarization measurement is indispensable, as firmly performed for cases of perovskite $R$MnO$_3$ and Ba$_2$Mg$_2$Fe$_{12}$O$_{22}$. The usefulness of which is also demonstrated in the study of a triangular antiferromagnet CuFeO$_2$. In the collinear spin ordered phase of CuFeO$_2$, two sharp resonances around 1.2 meV and  2.4 meV  in the transmission spectrum were already observed and assigned to AFMRs  based on the electron spin resonance (ESR) experiments \cite{TFukuda}. However, recent terahertz time-domain spectroscopic investigations on the basis of the light-polarization dependence was revealed that one of two resonances ($\sim2.4$ meV) can be assigned to the electromagnon \cite{SSeki}. 

Strong enhancement of versatile optical properties or ME effect in the dynamical regime via electromagnons is a big challenge to widen the bottle-neck of the spin-electronics technology. As such example, we showed the gigantic terahertz magnetochromism via electromagnons observed in Ba$_2$Mg$_2$Fe$_{12}$O$_{22}$. The microscopic origin of the electromagnon activity is relevant to the local polarization, which is modulated by the exchange striction mechanism. Therefore, the modification of the conical angle by external magnetic fields can lead to the gigantic terahertz magnetochromism as presented here. Another remarkable ME response via electromagnons is recently demonstrated in a square-lattice antiferromagnet Ba$_2$CoGe$_2$O$_7$ \cite{IKezsmarki}. By using terahertz time-domain spectroscopy, there found the noticeable two peak structures around 2 meV and 4 meV in the absorption spectrum. The former is ascribed to the AFMR driven by $H^\omega$, as previously revealed by the inelastic neutron scattering \cite{AZheludev}. However, the latter is assigned to the hybrid mode of the electric and magnetic excitations (or electromagnon) on the basis of the detailed light-polarization dependence. Noticeably, only near the electromagnon resonance, Ba$_2$CoGe$_2$O$_7$ shows the gigantic terahertz non-reciprocal directional dichroism \cite{IKezsmarki}, in which the light propagates in a different manner in forward and backward directions \cite{TArimaME}. Contrary to cases of $R$MnO$_3$ and Ba$_2$Mg$_2$Fe$_{12}$O$_{22}$, the observed non-reciprocal directional dichroism is argued within the framework of the spin-dependent metal-ligand hybridization mechanism \cite{SMiyaharaBCGO}.

An another challenging issue is the phase control via the coherent control of the electromagnon. As theoretically proposed in $R$MnO$_3$, the intense terahertz pulse with a peak amplitude in the order of MV/cm is possible to control the spin chirality ($S_i \times S_j$) and hence the ferroelectricity \cite{MMochizukiControl}. This would be the hallmark of the electromagnon as the medium of the versatile terahertz optical switch.

\section{Acknowledgments}

We thank the collaborators, especially, S. Kumakura, D. Okuyama, K. Iwasa, S. Ishiwata, Y. Taguchi, S. Miyahara, M. Mochizuki, T. Arima, for their enlightening discussions. This work was partly supported by Grants-In-Aid for Scientific Research (Grant No. 20340086 and 2010458) from the MEXT of Japan, and by FIRST Program by JSPS.






\begin{thebibliography}{00}


\bibitem{YTokuraREV} Y. Tokura, S. Seki, Adv. Mater. 22 (2010) 1554.

\bibitem{TArima} T. Arima, J. Phys. Soc. Jpn. 80 (2011) 052001.

\bibitem{HKatsura} H. Katsura, N. Nagaosa, A. V. Balatsky, Phys. Rev. Lett. 95 (2005) 057205.

\bibitem{IASergienko} I. A. Sergienko, E. Datotto, Phys. Rev. B 73 (2006) 094434.

\bibitem{Mostovoy} M. Mostovoy, Phys. Rev. Lett. 96 (2006) 067601.


\bibitem{TKimura} T. Kimura, T. Goto, H. Shintani, K. Ishizaka, T. Arima, Y. Tokura, Nature (London) 426 (2003) 55.





\bibitem{YTokuraNKida} Y. Tokura, N. Kida, Phil. Trans. Roy. Soc. A (2011) in press (Eprint arXiv:1104.0357).

\bibitem{PimenovREV} A. Pimenov, A. M. Shuvaev, A. A. Mukhin, A. Loidl, J. Phys.: Condens. Matter 20 (2008) 434209.

\bibitem{DSenffREV} D. Senff, N. Aliouane, D. N. Argyriou, A. Hiess, L. P. Regnault, P. Link, K. Hradil, Y. Sidis, M. Braden, J. Phys.: Condens. Matter 20 (2008) 434212.

\bibitem{NKidaRMnO3review} N. Kida, Y. Takahashi, J. S. Lee, R. Shimano, Y. Yamasaki, Y. Kaneko, S. Miyahara, N. Furukawa, T. Arima,  Y. Tokura,  J. Opt. Soc. Am. B 26 (2009) A35.

\bibitem{ABSushkov} A. B. Sushkov, R. Vald\'{e}s Aguilar, S. Park, S-W. Cheong, H. D. Drew, Phys. Rev. Lett. 98 (2007) 027202.

\bibitem{SPailhes} S. Pailh\`{e}s, X. Fabr\`{e}ges, L. P. R\'{e}gnault, L. Pinsard-Godart, I. Mirebeasu, F. Moussa, M. Hennion, S. Petit, Phys. Rev. B 79 (2009) 134409.

\bibitem{MCazayous} M. Cazayous, Y. Gallais, A. Sacuto, R. de Sousa, D. Lebeugle, D. Colson, Phys. Rev. Lett. 101 (2008) 037601.

\bibitem{NKidaYpart1} N. Kida, D. Okuyama, S. Ishiwata, Y. Taguchi, R. Shimano, K. Iwasa, T. Arima, Y. Tokura, Phys. Rev. B 80 (2009) 220406(R).

\bibitem{NKidaYpart2} N. Kida, S. Kumakura, S. Ishiwata, Y. Taguchi, Y. Tokura, Phys. Rev. B 83 (2011) 064422.

\bibitem{SSeki} S. Seki, N. Kida, S. Kumakura, R. Shimano, Y. Tokura, Phys. Rev. Lett. 105 (2010) 097207.

\bibitem{IKezsmarki} I. K\'{e}zsm\'{a}rki, N. Kida, H. Murakawa, S. Bord\'{a}cs, Y. Onose,  Y. Tokura, Phys. Rev. Lett. 106 (2011) 057403.

\bibitem{PDRogers} P. D. Rogers, Y. J. Choi, E. C. Standard, T. D. Kang, K. H. Ahn, A. Dubroka, P. Marsik, Ch. Wang, C. Bernhard, S. Park, S.-W. Cheong, M. Kotelyanskii, A. A. Sirenko, Phys. Rev. B 83 (2011) 174407.


\bibitem{RAguilar} R. Vald\'{e}s Aguilar, M. Mostovoy, A. B.  Sushkov, C. L. Zhang, Y. J. Choi, S-W. Cheong, H. D. Drew, Phys. Rev. Lett.  102 (2009) 047203.

\bibitem{SMiyahara} S. Miyahara, N. Furukawa, Eprint arXiv:0811.4082.

\bibitem{MPVStenberg} M. P. V. Stenberg, R. de Sousa, Phys. Rev. B 80 (2009) 094419.

\bibitem{MMochizuki_EM} M. Mochizuki, N. Furukawa, N. Nagaosa, Phys. Rev. Lett. 104 (2010) 177206.

\bibitem{APimenov1} A. Pimenov, A. A. Mukhin, V. Yu. Ivanov, V. D. Travkin, A. M. Balbashov, A. Loidl, Nat. Phys. 2 (2006) 97.

\bibitem{AAMukhin} A. A. Mukhin, V. Yu. Ivanov, V. D. Travkin, A. Pimenov, A. Loidl, A. M. Balbashov, Europhys. Lett. 49 (2000) 514.


\bibitem{NKidaDyMnO3} N. Kida, Y. Ikebe, Y. Takahashi, J. P. He, Y. Kaneko, Y. Yamasaki, R. Shimano, T. Arima, N. Nagaosa, Y. Tokura, Phys. Rev. B 78 (2008) 104414.


\bibitem{YTakahashi} Y. Takahashi, N. Kida, Y. Yamasaki, J. Fujioka, T. Arima, R. Shimano, S. Miyahara, M. Mochizuki, N. Furukawa, Y. Tokura, Phys. Rev. Lett. 101 (2008) 187201.

\bibitem{NKidaGdTbMnO3} N. Kida, Y. Yamasaki, R. Shimano, T. Arima, Y. Tokura, J. Phys. Soc. Jpn. 77 (2008) 123704.

\bibitem{JSLee} J. S. Lee, N. Kida, S. Miyahara, Y. Yamasaki, Y. Takahashi, R. Shimano, N. Furukawa, Y. Tokura, Phys. Re. B 79 (2009) 180403(R).

\bibitem{SIshiwataNeutron} S. Ishiwata, D. Okuyama, K. Kakurai, M. Nishi, Y. Taguchi, Y. Tokura, Phys. Rev. B 81 (2010) 174418.


\bibitem{SIshiwata} S. Ishiwata, Y. Taguchi, H. Murakawa, Y. Onose, Y. Tokura, Science 319 (2008) 1643. 

\bibitem{KTaniguchi} K. Taniguchi, N. Abe, S. Ohtani, H. Umetsu, T. Arima, Appl. Phys. Express 1 (2008) 031301. 

\bibitem{HSagayama} H. Sagayama, K. Taniguchi, N. Abe, T. Arima, Y. Nishikawa, S. Yano, Y. Kousaka, J. Akimitsu, M. Matsuura, K. Hirota, Phys. Rev. B 80 (2009) 180419(R).




\bibitem{NMomozawa} N. Momozawa, Y. Yamaguchi, M. Mita, J. Phys. Soc. Jpn. 55 (1986) 1350.


\bibitem{TFukuda} T. Fukuda, H. Nojiri, M. Motokawa, T. Asano, M. Mekata, Y. Ajiro, Physica B 246-247 (1998) 569.



\bibitem{AZheludev} A. Zheludev, T. Sato, T. Masuda, K. Uchinokura, G. Shirane, B. Roessli, Phys. Rev. B 68 (2003) 024428.

\bibitem{TArimaME} T. Arima, J. Phys.: Condens. Matter 20 (2008) 434211.

\bibitem{SMiyaharaBCGO} S. Miyahara, N. Furukawa, J. Phys. Soc. Jpn. in press (Eprint  arxiv:1101.3679).


\bibitem{MMochizukiControl} M. Mochizuki, N. Nagaosa, Phys. Rev. Lett. 105 (2010) 147202.




\end{thebibliography}



\begin{figure}[htbh]
\begin{center}
\includegraphics[width=0.85\textwidth]{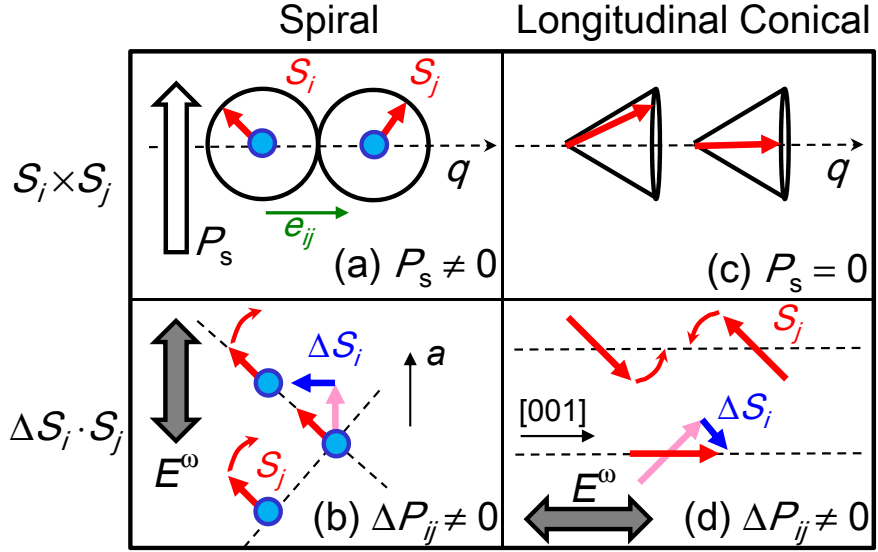}
\caption{(color online) Mechanisms of the spin-dependent polarization based on $S_i\times S_j$ and $\Delta S_i\cdot S_j$. (a) In spiral magnets such as $R$MnO$_3$, $P_{\rm s}$ (shown by an arrow) emerges according to Eq. (1), while (c) no $P_{\rm s}$ in the longitudinal conical spin ordered phase, as realized in Ba$_2$Mg$_2$Fe$_{12}$O$_{22}$. On the other hand, the exchange striction mechanism in the dynamical regime acts as a source of the electromagnon activity in both (b) spiral and (d) longitudinal conical magnets. In both cases, $\Delta P_{ij}$ becomes non-zero in response to the light $E$ field indicated by shaded arrows.}
\end{center}
\end{figure}

\begin{figure}[tb]
\begin{center}
\includegraphics[width=0.85\textwidth]{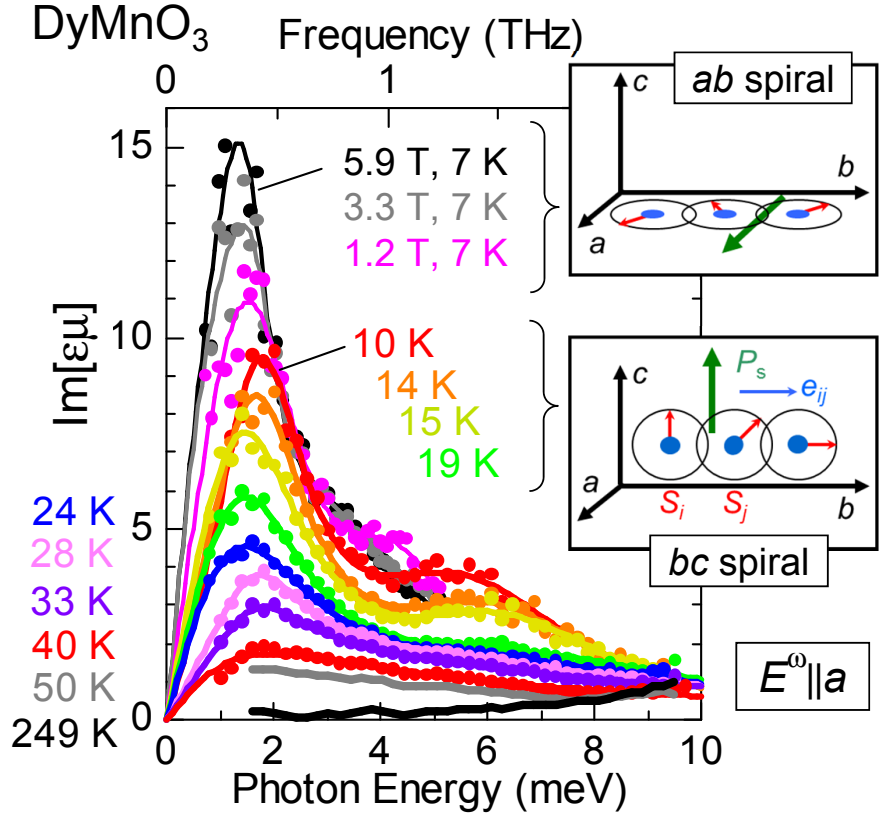}
\caption{(color online) Electromagnon in $bc$ and $ab$ spiral spin ordered phases of DyMnO$_3$ \cite{NKidaRMnO3review,NKidaDyMnO3}. Imaginary part of $\epsilon\mu$ spectra at selected temperatures and magnetic fields. The light $E$ vector was set parallel to $a$-axis. In the $bc$ spiral spin ordered phase below 19 K, two peak structures around 2 meV and 6 meV are discerned, which is a general characteristic of the electromagnon spectra in perovskite $R$MnO$_3$. $ab$ spiral spin ordered phase was induced by an application of the magnetic field along $b$-axis at 7 K.}
\end{center}
\end{figure}

\begin{figure}[tb]
\begin{center}
\includegraphics[width=0.85\textwidth]{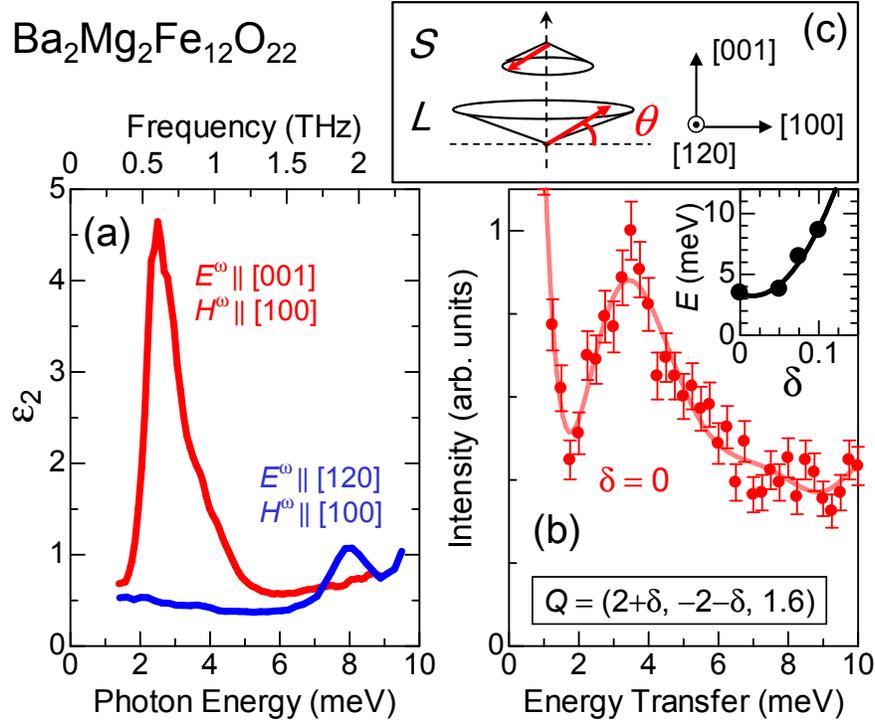}
\caption{(color online) Electromagnon in the longitudinal conical spin ordered phase of Ba$_2$Mg$_2$Fe$_{12}$O$_{22}$ \cite{NKidaYpart1}. (a) Imaginary part of $\epsilon$ spectra at 5 K, measured by using terahertz time-domain spectroscopy. Sharp resonance around 2.8 meV is visible when the light $E$ vector was set parallel to [001]. (b) Inelastic neutron scattering spectrum at zone center in the Brillouin zone, measured at 10 K. Inset shows the magnon dispersion with respect to $\delta$. (c) Schematic illustration of the longitudinal conical spin structure characterized by the conical angle $\theta$ with crystal orientation. }
\end{center}
\end{figure}

\begin{figure}[t]
\begin{center}
\includegraphics[width=0.61\textwidth]{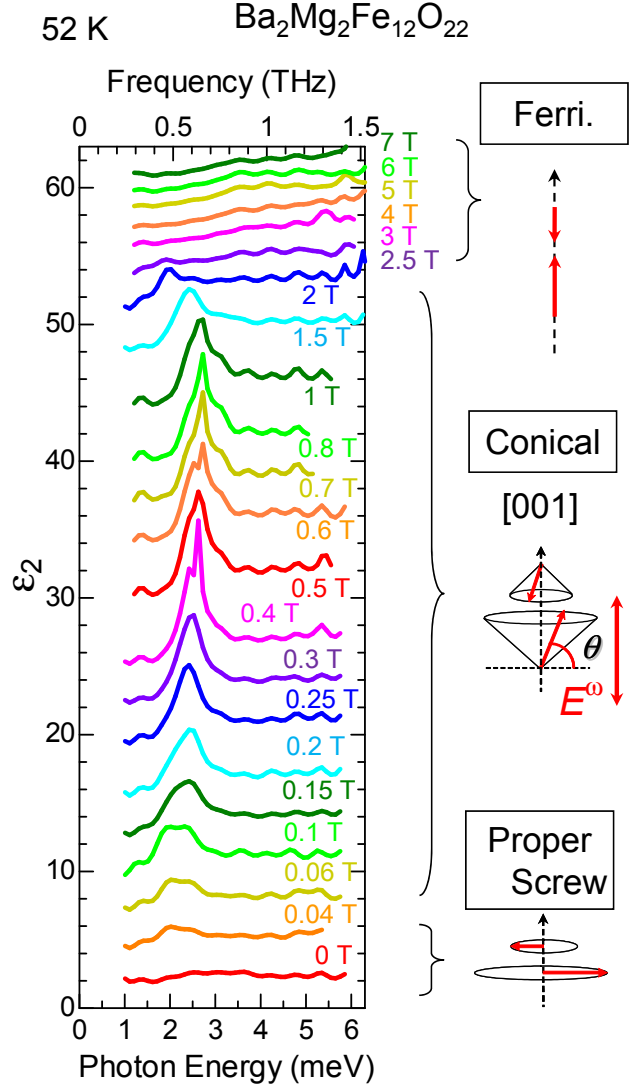}
\caption{(color online) Gigantic terahertz magnetochromism via electromagnons in Ba$_2$Mg$_2$Fe$_{12}$O$_{22}$ \cite{NKidaYpart2}. Imaginary part of $\epsilon$ spectra at selected magnetic fields up to 7 T, measured at 52 K; data are arbitrarily offset. Near 50 K, we can control the spin structures by applying the magnetic field along [001] from the proper screw (at 0 T) to the collinear ferrimagnetic ($>2.5$ T) through the longitudinal conical spin structures, as schematically shown in the right side of figure. As the electromagnon becomes active only in the longitudinal conical spin ordered phase, the spectral shape of the electromagnon dramatically changes in magnetic fields.}
\end{center}
\end{figure}

\end{document}